# Revealing plasma oscillation in THz spectrum from laser plasma of molecular jet

Na Li,[1] Ya Bai,[1] Tianshi Miao,[1] Peng Liu,[1,2,*] Ruxin Li[1,2] and Zhizhan Xu[1]

[1] *State Key Laboratory of High Field Laser Physics, Shanghai Institute of Optics and Fine Mechanics, Chinese Academy of Sciences, Shanghai 201800, China*
[2] *Collaborative Innovation Center of IFSA (CICIFSA), Shanghai Jiao Tong University, Shanghai 200240, China*

**Abstract:** Contribution of plasma oscillation to the broadband terahertz (THz) emission is revealed by interacting two-color ($\omega/2\omega$) laser pulses with a supersonic jet of nitrogen molecules. Temporal and spectral shifts of THz waves are observed as the plasma density varies. The former owes to the changing refractive index of the THz waves, and the latter correlates to the varying plasma frequency. Simulation of considering photocurrents, plasma oscillation and decaying plasma density explains the broadband THz spectrum and the varying THz spectrum. Plasma oscillation only contributes to THz waves at low plasma density owing to negligible plasma absorption. At the longer medium or higher density, the combining effects of plasma oscillation and absorption results in the observed low-frequency broadband THz spectra.

## 1. Introduction

Broadband ultrashort terahertz (THz) pulses are generated from the laser induced plasma by two-color pulses, and it is under intensive studies owing to the potential to produce high intensity THz field and achieve the off-site detection in air [1-7]. From the plasma filament pumped by the two-color laser pulses, the directional photocurrents generate the enhanced supercontinuum at the THz wavelength range with the conversion efficiency two orders larger than that from single color intense laser field [8,9]. The low-frequency broadband spectrum is determined by the constructive interference of the attosecond electron bursts whose timing is controllable by detuning the frequencies of the two-color pulses [10,11]. Based on the transient photocurrent model, schemes of controlling the THz polarization and spectrum have been proposed [11-15].

As the ultrashort THz pulses are produced from the ionizing gaseous medium, the surrounding plasma is expected to play an important role in the generation mechanism. The plasma oscillation has been considered as the origin of THz spectrum in previous theoretical studies on the generation of strong THz fields [16,17]. Recently A. Debayle *et al.* [18] proposed that the plasma current oscillation is in competition with the photocurrent mechanism and it can contribute to the THz spectrum when the length of medium is less than the skin depth of plasma. Also, the residual plasma oscillation after the interaction of the pumping laser pulses, instead of the photocurrents when the laser is on, was proposed as a source of the low-frequency THz emission [19-21]. However, experimental investigations showed no evidence of plasma oscillation in play, possibly hindered by the complex dynamics of laser propagation in filaments, and how the laser plasma impacts the broadband THz generation remains controversial.

We experimentally investigate the THz generation from a jet of nitrogen ($N_2$) molecules pumped with the two-color laser pulses. The molecular beam in vacuum provides an medium of limited transmitting length to observe the THz waves under the conditions of varied plasma density. We observed frequency shift at the low plasma density and found that the changed THz frequency can only correlate to the varied plasma frequencies. The observation suggests that the plasma oscillation contributes to THz spectrum only at low plasma density owing to negligible plasma absorption when the medium length is less than the plasma depth. Our results also show that as the plasma density increases the spectrum of THz approaches to a constant central frequency, indicating a more dorminant role of the absorption of plasma. In simulation we consider the decay rate of plasma density as an amendment to the semi-analytical one-dimensional model [18].

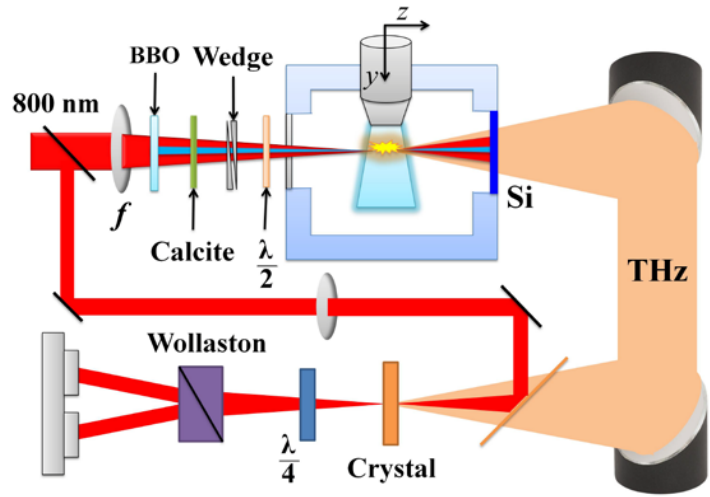

Fig. 1. Schematic diagram of the experimental setup.

## 2. Experimental implementation and results

Experimentally we prepare a short length of nitrogen ($N_2$) molecules using the molecular beam method [22]. A beam of $N_2$ molecules is produced from supersonic expansion of a nozzle (General Valve Series 9, orifice diameter of 0.5 mm) in a high vacuum chamber ($10^{-4}$ Pa). The jet is mounted on a mortorized stage and its position is adjustable with the minimum step size of 0.05 mm along the directions of laser propagation (*z*-axis) and jet expansion (*y*-axis). Laser pulses are delivered along the *z* direction from a commercial Ti: sapphire laser amplifier (Coherent Elite-HP-USX) with the duration of 40 fs and repetition rate of 1 kHz at the center wavelength of 800 nm. As shown schematically in Fig. 1, a portion of the laser pulses (1.2 mJ) passes through a focusing lens (*f* = 300 mm), a 200-$\mu$m-thick type I beta barium borate (β-BBO) crystal, a calcite for compensating the group delay of the two color pulses, and a zero-order λ/2 (400nm) waveplate for rotating the polarization of the generated 400 nm pulses to be parallel with the 800 nm pulses. The phase delay of the two color pulses is controlled with a pair of fused silica wedges and the THz yield modulates as the phase delay changes [23]. The two-color pulses focuses onto the molecular beam at 0.5 mm downstream of the orifice, where the interaction length is estimated to be $L_m$ = 0.7 mm by the fluorescence of the nitrogen cations in the plasma. The left of the laser pulses is used as the probe pulses for the measurement of time-domain THz waveforms using the electro-optic (EO) sampling method [24]. For the EO sampling, both a ZnTe crystal of 100 $\mu$m thickness and a GaP crystal of 200 $\mu$m are used, allowing the cutoff of the measured bandwidth to ~8 THz. The response function of the crystals are calculated to correct the measured THz waves.

By altering the stagnation pressure of the jet, we achieve the experimental condition where all the experimental parameters remain constant except for the number density of molecules at the focus. Fig. 2(a) shows that the obtained THz waveforms vary significantly as the pressure changes from 0 to 1.0 bar. The THz waves at 0.2 bar and 0.7 bar are retrieved and the Wigner-transformed maps are plotted in Fig. 2(b) and 2(c) respectively. One can see that the center position of the spectra shifts from 0.8 THz to 1.2 THz while the temporal envelope from 5.2 ps to 5.0 ps. It is noted that, when using the OE sampling crystal of GaP (detection range 0.1~8 THz), the main body of THz spectrum show is still in low frequency range of 0~3 THz, and it represents the same feature of frequency shift.

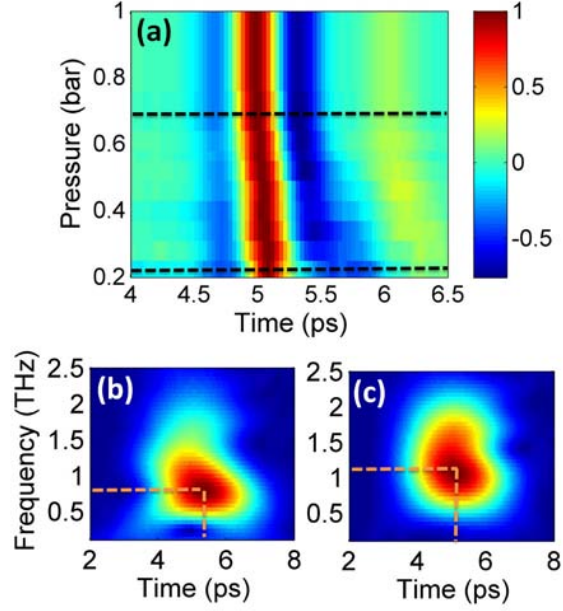

Fig. 2. (a) Normalized THz waveforms as a function of stagnation pressure of molecular jet (100 μm ZnTe crystal), and the time-frequency analysis of THz waves at 0.2 bar (b) and 0.7 bar (c).

## 3. Discussions

The temporal shift of the THz envelope can be attributed to the variation of the refractive index for the pumping laser pulses and the generated THz waves in the plasma, which is mainly determined by the varied gas densities. The temporal positions of the THz waves shown in Fig. 2(a) are calculated and plotted in Fig. 3(b), which shows an overall shift of ~ 0.2 ps. It is known that the frequency-dependent refractive index of the $N_2$ molecules composes of the two contributions: of the neutral molecules and of the plasma [25]

$$n(\omega) = 1 + \delta n_{gas}(\omega) + \delta n_{plasma}(\omega), \tag{1}$$

where $\delta n_{gas}(\omega)$ is the refractive index drift caused by neutral molecules and $\delta n_{plasma}(\omega)$ by plasma. We calculate the $\delta n_{gas}$ for THz waves by extrapolating the dispersion formula [25,26]. For the plasma drift, we use $n_{plasma}(\omega) = (1-\omega_p^2/\omega^2)^{1/2}$ for the pumping laser pulses and $n_{plasma}(\omega) = (1-\omega_p^2/(\omega^2+iv\omega))^{1/2}$ for the THz because the considered wavelength is close to the collision frequency [8]. Under the experimental conditions, the density of $N_2$ molecules and the plasma frequency are calculated by the empirical formula of the free jet expansion [22] and the Ammosov-Delone-Krainov (ADK) theory for ionization rate [27], respectively.

The variations of the refractive indexes for 800 nm pump pulses and the THz waves (1 THz) are calculated and shown in Fig. 3(a), respectively. One can see that the change of the refractive index for 800 nm is 4 orders lower than that for the THz waves. Calculation also indicates that the refractive index drift caused by plasma $\delta n_{plasma}$ is 9 orders larger than by neutral molecules $\delta n_{gas}$, so the velocity change of THz waves by plasma dominates the temporal envelope position. Fig. 3(b) also shows the calculated time shift of the THz waves by $\Delta t = Ln/c$ with $L$ the interaction length of 0.7 mm, which is consistent with the experimental observation. During the laser pulse of tens of femtoseconds, electrons are tunneling ionized and the time scale of THz emission extends to picoseconds. So the THz radiation is inevitably affected by surrounding plasma environment. Since the temporal advance of the THz waves is sensitively related to the plasma density, the phenomenon offers a means of measuring the plasma density *in situ* accurately.

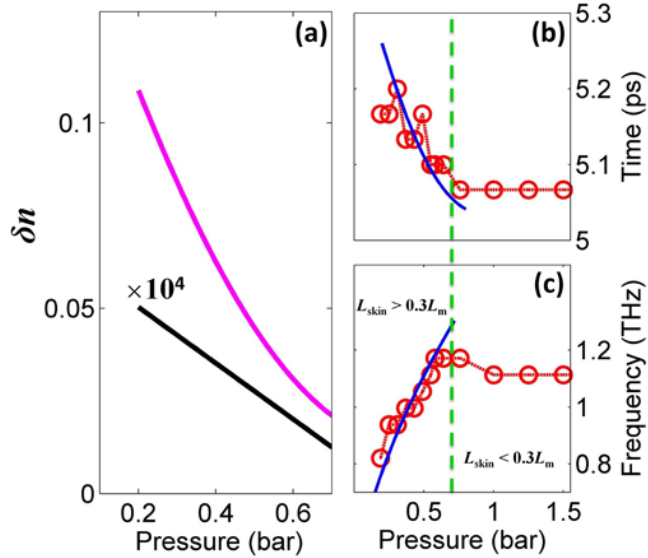

Fig. 3. (a) Calculated refractive index for 800 nm (black line) and THz waves (1 THz, purple line) at varied stagnation pressures, respectively; (b) the center positions (red solid circles) of the temporal envelope of THz waves and simulated temporal advance (blue solid line) of the THz pulses (~1 THz); (c) the THz frequencies (red solid circles) and the calculated 0.7 $f_p$ (blue solid line). Note that for the pressure less than 0.7 bar (green dotted line), $L_{skin} > 0.3 L_m$, the THz frequencies is sensitive to the plasma environment.

With the varying plasma density, the underlying physics of the spectral shift can be analyzed. We consider the two possible mechanisms of the frequency shift during the generation of transient photocurrents: one is the spatially varied phase difference of the two-color laser pulses, $\varphi(z)$, in the plasma; and the other is originated from the detuning of spectrum of the driving laser pulses. For the former case, the dispersion within the short length of plasma increases the phase difference of the two color pulses, which can modify not only the generated THz spectrum but also the yield [5,8]. We calculate the increased phase difference $\delta\varphi = \varphi(0) - \varphi(L_m) = 0.0086\,\pi$ for 0.2 bar and $\delta\varphi = 0.038\,\pi$ for 1 bar, which are little to modulate the THz yield. Secondly, the detuning of the two-color laser pulse spectrum in plasma may change the timing of the attosecond electron bursts and shift the generated THz frequency through the interference [11,15]. We calculate the shifted spectrum of the driving two-color laser pulses under the varying refractive index induced by laser plasma at the pressure of 0.2 bar and 0.7 bar (the blue shift for 800nm is 0.1 THz and for 400nm 0.05 THz), and obtain the THz field in frequency domain by $E_{\text{THz}}(\omega) \propto \omega J(\omega) = B(\omega)\sum C_n e^{i\omega t_n}$, where $C_n$ is relative to the step-wise increase of the electron density and $B(\omega) = \frac{\omega}{\omega - i\nu}\exp(-\frac{\omega^2}{\sigma^2})$, in which $\nu$ is the phenomenological electron-ion collision rate, $\sigma \approx 10$ fs$^{-1}$ [11]. As a result, the calculated frequency shift of THz is less than 0.01 THz, much less than the experimental results of 0.4 THz as shown in Fig. 3(b). Based on the shifted spectrum of lase pulses, we also calculate the THz spectrum using the rectification by four-wave mixing (Kerr or bound electron response) [14,28], the THz frequency shifts ~0.0045 THz, which is in a difference of an order with observed THz frequency shift.

The discrepancy of the above simulation with the experimental observation indicates the contribution of plasma frequency to THz spectrum, as the small-size beam and low stagnation pressure of molecules permits a medium length that is shorter than the skin depth of plasma [8,18]. For an uniformly distributed plasma in a cylinder volume, the resonance frequency is $f_c = f_p/\sqrt{2} = 0.707 f_p$. The peak frequency of THz spectrum is approximately equal to the plasma resonance frequency, namely $f_{\text{THz}} = f_c$ [20]. We plot the observed THz frequencies and the calculated frequencies of $f_c$ in Fig. 3(c), and the two curves match well. The consistence indicates that the plasma oscillation plays a major role in the obtained THz spectra.

Notably, we observe that the THz central frequency varies under low stagnation pressures and converges to a constant central frequency at around 0.7 bar, where the plasma skin depth is close to the medium length, $L_{\text{skin}} = c/[\omega \text{ Im}(n)] = 0.3\ L_{\text{m}}$. Under the experimental condition of the changing gas pressure, the saturation of electron density can be ruled out [29]. It has been proposed that the plasma current oscillation can mainly contribute to the THz spectrum when plasma skin depth is larger than the medium length, and the THz peak frequency increases with plasma frequency for $L_{\text{skin}} > L_{\text{m}}$ [18]. The local electric field $\delta E$ emitted by the plasma is expressed as

$$\delta E \approx \frac{\exp(-\nu t/2)}{\sqrt{\omega_{\text{cf}}^2 - \frac{\nu^2}{4}}} \sin\left(-\sqrt{\omega_{\text{cf}}^2 - \frac{\nu^2}{4}}\, t\right) \cdot G, \qquad (2)$$

where $G = \frac{e^2}{m_e} \int_0^{\min(2\tau,t)} n_e(t')E_L(t')dt'$, $v$ is the electron-ion collisional frequency, $\omega_{cf}$ is the calculated $2\pi f_c$ under the final electron density ($n_{ef}$) after gas ionization, $E_L$ is the laser electric field, and $\tau$ is the laser pulse width. So the THz field oscillates at plasma resonance frequency. When $L_{skin} < L_m$ at higher plasma density, the photocurrent mechanism becomes dominant, and the THz peak frequency ought to corresponds to the plasma frequency, shown in Fig. 7 of Ref. [18]. However, our experimental result indicates otherwise for $L_{skin} < 0.3 L_m$.

The converging of THz spectrum to a constant central frequency can be explained by introducing the decay rate of plasma density into the semi-analytical one-dimensional model [18]. In the laser plasma generated from the molecular jet, the time-evolution of electron density can be expressed as [30,31]

$$n_{eft}(t) = n_{ef} / (1 + \beta n_{ef} t), \tag{3}$$

where $\beta$ is the decay coefficient used to characterize the electron-ion recombination. The decay rate $\beta$ has been measured to be $\beta = 3.1 \times 10^{-12}$ $m^3/s$ in the previous study of Ref. [32]. The ratio between $L_{skin}$ and $L_m$ affects the THz spectrum mainly by absorption during transmission in plasma. The absorption function of THz spectrum by plasma [8] is

$$F(\omega, z) \propto \exp(-(L_m - z)/L_{skin}(\omega)), \tag{4}$$

where $z$ is the position of THz generation. The THz spectrum is then obtained by integrating the THz emission along the plasma spatial dimension [33]

$$E_{THz}(\omega) \propto \int_0^{L_m} \pi w^2(z) F(\omega, z) \delta \hat{E}(z, \omega) \exp(i\theta(z)) dz, \tag{5}$$

where, $w(z)$ is the beam radius at different position, $\delta \hat{E}(z, \omega) = \int_{-\infty}^{\infty} \delta E(z,t) e^{-i\omega t} dt$, $\theta(z)$ is the phase at the exit of the plasma. The simulated results of $E_{THz}(\omega)$ is shown in Fig. 4, and it is consistent with experimental results regarding the turning to a constant central frequency.

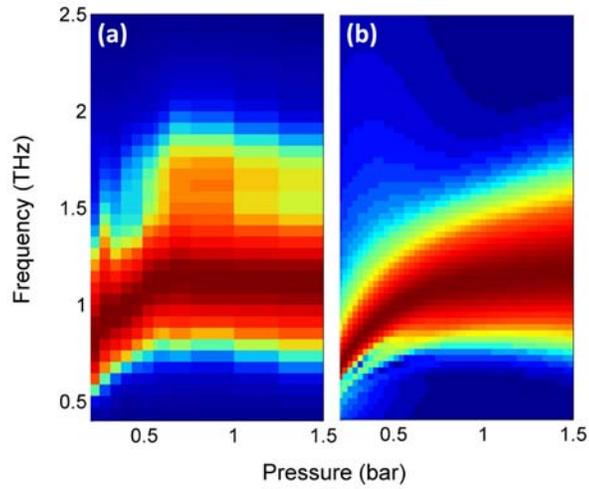

Fig. 4. (a) Normalized THz spectrum as a function of stagnation pressure; (b) simulated THz spectrum.

The longer skin depth for low plasma density leads to negligible absorption according to Eq. (4). Hence the contribution of plasma oscillation can be observed experimentally in Fig. 3(c) for low pressure. On the contrary, the shorter skin depth for high plasma density leads to the generated $\delta E$ strongly absorbed in the plasma. So the peak frequency of THz no longer change under the varied dense plasma density, but with decay of plasma density after ionization, the skin depth will change longer until that the skin depth become close to the length of the plasma, in which $\delta E$ with the plasma oscillation characteristics is reappear. So the transmitted plasma resonance frequency under high plasma density is directly related to the medium length but not the plasma density and keep invariant.

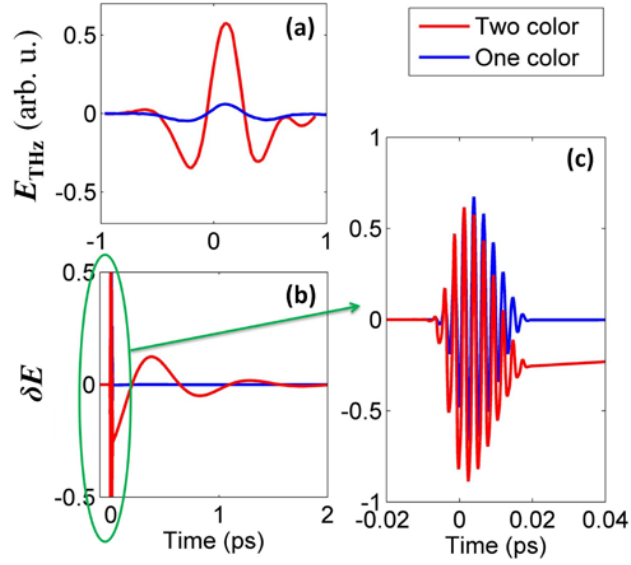

Fig. 5. (a) Measured terahertz electric field waveform for two-color pulse (red line) and one-color pulse (blue line), respectively. (b) Simulated unfiltered local $\delta E(t)$ for two-color pulse (red line) and one-color pulse (red line). (c) $\delta E(t)$ is enlarged in time scale of -0.02 ps~0.04 ps of Fig. 5(b)

Experimentally we also found that the amplitude of the THz waves from the jet molecules decreases 10 times as the driving laser field changes from two color pulses to one color pulse), shown in Fig. 5(a). The last item G in Eq. (2) indicates a non-zero DC electron current density after the laser pulse. One can see that two color pulses can effectively increase G, as shown in Fig. 5(c), and the THz yield from plasma oscillation also increases effectively as shown in Fig. 5(b). It explain well why the amplitude of the THz waves by one color driving pulses decreases much even through it is contributed from plasma oscillation.

There are still some errors in our simulation result comparing to the experiments shown in Fig. 3(b). We propose two possible reasons that may improve the model to explain the results. Firstly, the increase of plasma density may lead to the generated off-axis phase matching THz and reduce the plasma influence in the transmission. Because it start to make the transition to photocurrent mechanism [7,14] (conical THz radiation has been found in related studies [34,35] based on photocurrent in long plasma filament). Secondly, under the high back pressure, the generated THz wave may experiences a more complex propagation effect such as the plasma defocusing effect and the diffraction, with which one may improve the accuracy of simulation.

## 4. Summary

In conclusion, identification of the contribution of plasma oscillation is the first time for the generation of broadband THz pulses in ultrashort two-color laser plasma in experiments. Temporal and spectral shifts of THz waves are observed as the plasma density varies at low backing pressure. The former owes to the changing refractive index of the THz waves, and the latter correlates to the varying plasma frequency. Simulation of considering photocurrents, plasma oscillation and decaying plasma density explains the broadband THz spectrum and the varying THz spectrum. Plasma oscillation only contributes to THz waves at low plasma density owing to negligible plasma absorption. At the longer medium or higher density, the combining effects of plasma oscillation and absorption results in the observed low-frequency broadband THz spectra. Our study provides significant guidance for the tunable terahertz spectrum radiation based on plasma oscillation.

## Acknowledgment

The authors acknowledge the support of the Chinese Academy of Science, the Chinese Ministry of Science and Technology, the National Science Foundation of China (Grant Nos. 11274326, 61521093, 61405222, 11134010 and 11127901), Shanghai Sailing Program (Grant No. 14YF1406200) and the 973 Program of China (2011CB808103).